\documentclass[twocolumn,noshowpacs,amsmath,amssymb,prl]{revtex4}

\usepackage{graphicx}
\usepackage{amsmath}

\newcommand{\beginsupplement}{%
        \setcounter{table}{0}
        \renewcommand{\thetable}{S\arabic{table}}%
        \setcounter{figure}{0}
        \renewcommand{\thefigure}{S\arabic{figure}}%
     }

\begin{document}

\title{Overcoming power broadening of the quantum dot emission \\in a pure wurtzite nanowire}



\author{M.E. Reimer$^*$, G. Bulgarini, R.W. Heeres, B.J. Witek, M.A.M. Versteegh, and V. Zwiller}
 \email{m.e.reimer@tudelft.nl}
 \affiliation{Kavli Institute of Nanoscience, Technical University of Delft, Delft, The Netherlands}

\author{D. Dalacu, J. Lapointe and P.J. Poole}%
\affiliation{National Research Council of Canada, Ottawa, Canada}


\begin{abstract}
One of the key challenges in developing quantum networks is to generate single photons with high brightness, purity, and long temporal coherence. Semiconductor quantum dots potentially satisfy these requirements; however, due to imperfections in the surrounding material, the coherence generally degrades with increasing excitation power yielding a broader emission spectrum. Here we overcome this power broadening regime and demonstrate the longest coherence at exciton saturation where the detection count rates are highest. We detect single-photon count rates of 460,000 counts per second under pulsed laser excitation while maintaining a single-photon purity of greater than 99\,$\%$. Importantly, the enhanced coherence is attained with quantum dots in ultraclean wurtzite InP nanowires, where the surrounding charge traps are filled by exciting above the wurtzite InP nanowire bandgap. By raising the excitation intensity, the number of possible charge configurations in the quantum dot environment is reduced, resulting in a narrower emission spectrum. Via Monte Carlo simulations we explain the observed narrowing of the emission spectrum with increasing power. Cooling down the sample to 300\,mK, we further enhance the single-photon coherence two-fold as compared to operation at 4.5\,K, resulting in a homogeneous coherence time, $T_2$, of 1.2\,ns.
\end{abstract}

\maketitle

A quantum network is constituted by local nodes where quantum information is generated, processed, and stored, as well as communication channels between these nodes to coherently transfer quantum states across the entire network \cite{Kimble08}. The messengers of choice to distribute quantum information over long distances are single photons since they interact very weakly with the environment, thereby preserving their coherence, and are compatible with existing telecommunication fiber technologies. Coherent single photons are necessary in future quantum technologies such as linear optics quantum computing \cite{Knill01}, quantum teleportation \cite{Fattal04}, interfacing remote quantum bits \cite{Hofmann12, Bernien13} or integration of a quantum repeater \cite{Briegel98}. In addition to coherence, the source brightness and single-photon purity (i.e., suppressed multiphoton emission) are extremely important toward practical implementation of these quantum technologies.

Semiconductor quantum dots embedded in photonic structures are leading candidates to generate coherent and bright sources of single photons with high purity. However, increasing the source brightness typically comes at the cost of degrading the coherence resulting in a broadened emission spectrum \cite{Gerard06, Ates09, He13, Gazzano13}. This effect, known as `power-broadening', is attributed to an increase in the charge fluctuations of the quantum dot environment that leads to spectral wandering \cite{Gerard06}. Here, we report on the single-photon coherence from an ultra-clean material system comprising of a single quantum dot in a pure wurtzite InP nanowire, where dephasing mechanisms are suppressed by both cooling and higher excitation power. Remarkably, in stark contrast to conventional self-assembled quantum dots \cite{Gerard06,Ates09,He13,Gazzano13,PowBr_Houel,PowBr_Robinson,PowBr_Shih,PowBr_Stufler,PowBr_Ware}, we obtain the longest coherence at the highest brightness of the single-photon emission. Importantly, we detect single-photon count rates of $\sim$460 kilocounts per second while maintaining a single-photon purity of greater than 99\,$\%$. This latter feature provides a significant advantage over cavity based nanostructures where increased multiphoton events occur at saturation of the quantum dot emission due to cavity feeding of other quantum dots or detuned transitions of the same quantum dot \cite{Suff09,Winger09}. Finally, by cooling the sample to 300\,mK we show that the coherence can be enhanced by a factor of $\sim2$ as compared to operation at 4.5\,K by further suppressing interactions with phonons.
\begin{figure*}[t]
\centering
\includegraphics[width=14cm]{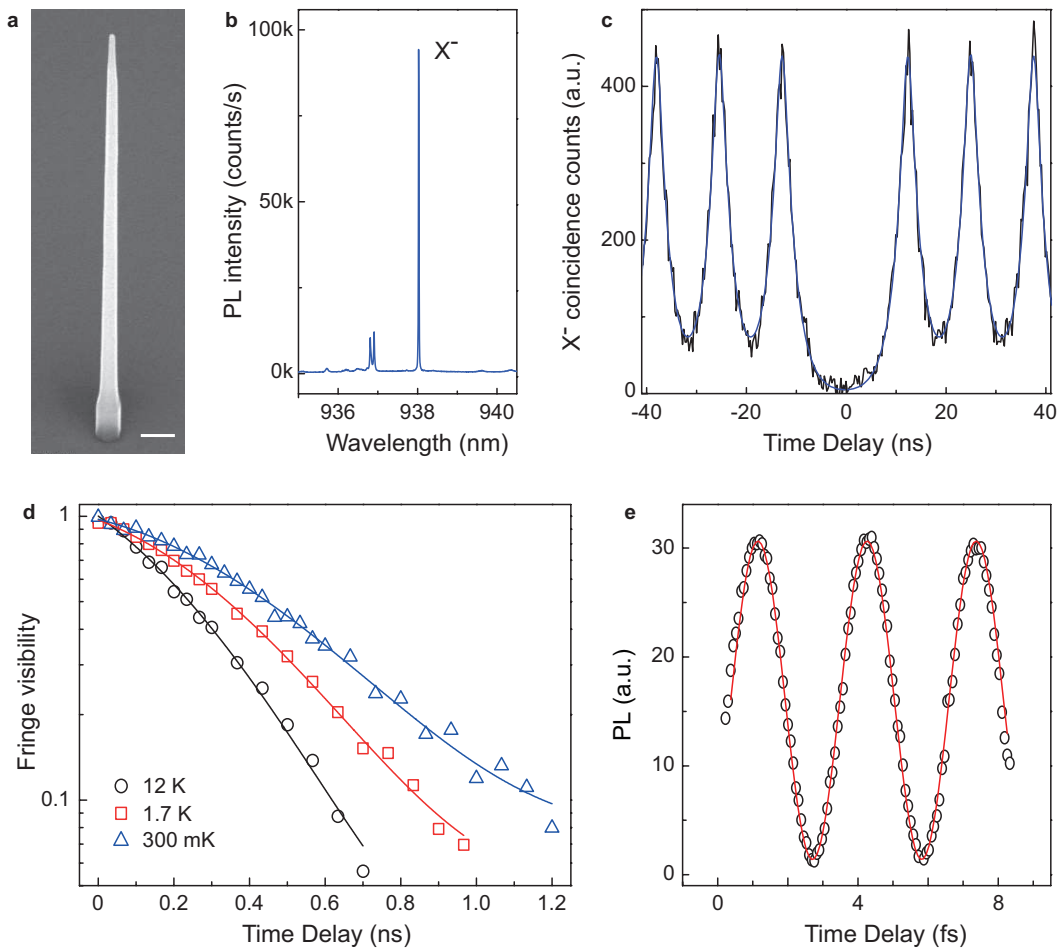}
\caption{\label{fig:1}\textbf{Single-photon interference measurements.} (a) SEM image of tapered nanowire waveguide containing a single quantum dot. Scale bar: 200\,nm. (b) Typical PL spectrum of a quantum dot that is predominantly charged. $X^-$: charged exciton. (c) Autocorrelation measurement of $X^-$ demonstrating a single-photon purity of greater than 99\,$\%$. A fit to the data (blue line) yields $g^2(0)<0.008$. (d) Single-photon interference measurements of a single quantum dot at 300\,mK (blue triangles), 1.7\,K (red squares), and 12\,K (black circles). The solid lines are a fit to the data using a Voigt profile (described in the text). (e) Raw single-photon interference fringes with a step size of 20 nm at maximum fringe visibility for a temperature of 300\,mK.}
\end{figure*}

A scanning electron microscope image of a typical InP tapered nanowire waveguide containing a single InAsP quantum dot used in the present experiments is shown in Fig.~\ref{fig:1}(a). There are three main features to this design in order to achieve both bright single-photon emission and a long temporal coherence. First, a waveguide shell is grown around the quantum dot with a very small taper towards the tip ($\sim1\,^{\circ}$) to boost single-photon collection efficiency by guiding the light efficiently towards the collection optics \cite{Reimer12}. Second, by design, the single quantum dot is always on the nanowire waveguide axis, which is needed for efficient coupling of the dipole emitter to the fundamental waveguide mode \cite{Bulgarini12}. Finally, the nanowire consists of a pure wurtzite InP crystal structure that is free of stacking faults. Importantly, this crystal phase purity of the nanowire waveguide is obtained for core diameters of 25\,nm or less and is maintained during shell growth to construct the waveguide \cite{Dalacu12}. The crystal phase purity is essential in obtaining a narrow emission spectrum: the measured linewidths are reduced by more than two orders of magnitude as compared to when stacking faults were present in nanowires \cite{Skold_PRB09,Heinrich10,Dal11,Reimer12}. Typically,  the presence of stacking faults in the nanowire act as efficient charge traps, which is detrimental to the quantum dot emission linewidth owed to spectral wandering.

As an added advantage, the nanowire heterostructures presented in this work are also deterministically positioned by combining both electron beam patterning and selective area epitaxy \cite{Dal11,Dalacu12}. The main important features pertaining to the single-photon quality that we demonstrate are long coherence, high light collection efficiency, and suppressed multi-photon emission. These main attributes when combined together have not been achieved to date for site-controlled quantum dots \cite{Reimer09,Dalacu10,Heinrich10,Bulgarini12,Schneider12,Jons13}.

\section{Single-photon coherence measurements}
To determine the coherence length of single photons originating from quantum dots in nanowire waveguides, and obtain a high resolution measurement of the emission spectrum, we employ field-correlation measurements using a Michelson interferometer \cite{Kam02,Zwiller04,Clemens12}. To ensure good overlap of the spatial modes between both paths, the quantum dot emission is coupled to a single mode fiber before entering the Michelson interferometer. When the path difference between both paths is varied we observe single-photon interference through oscillations in the measured intensity at the output, where the decay of the interference fringe visibility is set by the temporal coherence of the source.

The photoluminescence spectrum of the quantum dot used for the field-correlation measurements is presented in Fig.~\ref{fig:1}(b). The brightest line for this particular quantum dot at 938 nm is the singly charged exciton, X$^{-}$. Assignment of X$^{-}$ is confirmed by a combination of both polarization analysis as in Ref. \cite{Dalacu12} and cross-correlation measurements (see Supplementary Information). At saturation of the quantum dot emission, we measured count rates of 460 kilocounts per second on a silicon avalanche photodiode (APD) under pulsed excitation. Importantly, this high measured count rate is achieved while maintaining the single-photon purity of greater than 99\,$\%$ for the source, see Fig.~\ref{fig:1}(c). We excite the dot above the nanowire bandgap at 750\,nm using 3\,ps laser pulses at a repetition rate of 76\,MHz.  Accounting for a transmission of 1.4(1)$\%$ in the optical setup, we calculate a single-photon source efficiency of 43(4)$\%$. The single-photon source efficiency can be further enhanced by integrating a bottom gold mirror with a thin dielectric at the base in order to collect the downward emitted photons that are lost \cite{Claudon10,Reimer12}.

The results of the single-photon interference measurements are presented in Fig.~\ref{fig:1}(d) for three different temperatures from 12\,K down to 300\,mK. Fig.~\ref{fig:1}(e) depicts the raw interference fringes of the charged exciton line at 300\,mK obtained around zero path difference between both arms of the Michelson interferometer for a step size of 20\,nm. A fit to the data (red line) yields a maximum raw fringe visibility of 95\,$\%$. Correcting for the Michelson system response, a maximum fringe visibility of 99$\%$ is attained (see Methods). For increasing path difference (time delay), a decay in the fringe visibility is observed and we extract the coherence length, and corresponding coherence time, of the emitted single photons.

\begin{figure}[t]
\centering
\includegraphics[width=1.0\linewidth]{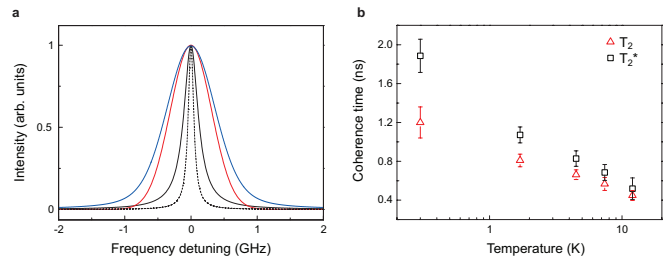}
\caption{\label{fig:2} \textbf{Single quantum dot coherence.} (a) Single quantum dot emission linewidth is represented by a Voigt profile (blue) with full-width half-maximum of 880(130)\,MHz, which is a convolution of a Lorentzian (black) and Gaussian (red) lineshape. The parameters for the homogeneous broadening (Lorentzian, $T_2=1.2(2)$\,ns, corresponding to 260(40)\,MHz) and inhomogeneous broadening (Gaussian, $T_c=0.9(1)$\,ns, corresponding to 730(50)\,MHz) are extracted from the fit of the Michelson data at 300\,mK in Fig. 1(d). For comparison, the dotted black line shows the lifetime Fourier-transform limit of 100\,MHz. (b) Coherence time, $T_2$, and corresponding pure dephasing time, $T_2^*$, extracted from the fits of Fig. 1(d). At 300\,mK the pure dephasing time is longer than the exciton lifetime, $T_1$, of 1.6(1)\,ns.}
\end{figure}

We now discuss the nature of the lineshape that is obtained from the single-photon interference measurements in Fig.~\ref{fig:1}(d). We observe a Voigt profile, which is a Gaussian convolved with a Lorentzian. Gaussian line broadening is due to charge fluctuations in the quantum dot environment, leading to spectral wandering on time scales longer than the exciton lifetime, but shorter than the measurement time needed to acquire a visibility data point ($\sim$5 minutes in our measurements). From a fit to the data (open symbols) at 300\,mK, we extract a homogeneous (Lorentzian) linewidth of 260(40)\,MHz and inhomogeneous (Gaussian) linewidth of 730(50)\,MHz (see Methods). Both components of the Voigt profile are plotted in Fig.~\ref{fig:2}(a), represented by the black (Lorentzian) and red (Gaussian) line. The convolution of both lineshapes yields the measured Voigt profile with a full-width half-maximum of 880(130)\,MHz, shown in blue. For comparison the dotted black line shows the lifetime Fourier-transform limit of 100\,MHz, for our measured exciton lifetime, $T_1$, of 1.6(1)\,ns (see Supplementary Information). The deviation from the Fourier-transform limit is mainly due to Gaussian line broadening. We point out that this Gaussian line broadening mechanism due to charge fluctuations in the environment is also observed in integrated resonant fluorescence \cite{Ates09, He13}, and found to degrade the visibility of two-photon interference from independent quantum dots \cite{Flagg10,Patel10,Wurzburg14}.

\section{Counteracting dephasing mechanisms}
In Fig.~\ref{fig:2}(b), we present the homogeneous coherence time, $T_2$, as extracted from the Voigt profile and corresponding pure dephasing time, $T_2^*$, from the relation
\begin{equation}
\frac{1}{T_2}=\frac{1}{2T_1}+\frac{1}{T^*_{2}},
\end{equation}
as a function of temperature. Here, $T^*_{2}$ is the pure dephasing, and $T_2$ is the homogeneous coherence time that includes both relaxation and pure dephasing processes. Decreasing the temperature manifests into an enhancement in the single-photon coherence as interactions with phonons are suppressed. Both $T_2$ and $T^*_{2}$ show an exponential increase while cooling. Interestingly, the homogeneous coherence time doubles to 1.2(2)\,ns when cooling from 4.5\,K to 300\,mK, while the pure dephasing time surpasses the measured exciton lifetime at 300\,mK. The corresponding inhomogeneously broadened linewidth reduces at a much weaker rate from 1.42\,GHz at 12\,K to 880\,MHz at 300\,mK (see Supplementary Information). The observed reduction in emission linewidth with temperature is consistent with previous work where quantum dot emission linewidths were reduced by as much as 120\,MHz/K \cite{Borri01, Bayer02}.

\begin{figure}[t]
\centering
\includegraphics[width=1.0\linewidth]{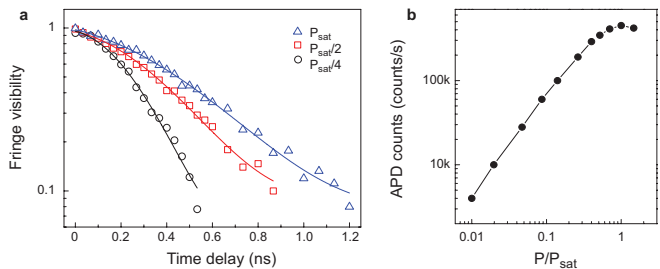}
\caption{\label{fig:3} \textbf{Power narrowing of the single quantum dot emission linewidth.} (a) Single-photon interference measurements of a single quantum dot at 300\,mK as a function of excitation power, P. In contrast to conventional self-assembled quantum dots, importantly, the best coherence is observed at exciton saturation of the quantum dot emission where the detected count rate is highest. The laser power, P$_{sat}$, used to saturate the exciton transition is $\sim$ 3\,$\mu$W (spot size of $\sim$ 1\,$\mu$m). (b) Power dependence of charged exciton emission line. The exciton emission saturates at 460,000 counts per second as measured on a single-photon avalanche photodiode.}
\end{figure}

Finally, we present in Fig.~\ref{fig:3}(a) the single-photon interference measurements at 300\,mK as a function of excitation power.  Remarkably, we observe the longest coherence at saturation of the quantum dot emission where the measured count rates on the single-photon APD are highest (see Fig.~\ref{fig:3}(b)). The resulting emission linewidth, to be discussed subsequently, is reduced from 2.7(3)\,GHz at the lowest excitation power used to 0.8(2)\,GHz at saturation of the quantum dot emission (see Fig.~4(d)). Excellent fits to the data in Fig.~\ref{fig:3}(a) are made by fixing $T_2$ to 1.2(2)\,ns at the saturation power, $P_{sat}$, of the quantum dot emission and increasing the degree of Gaussian line broadening as the excitation power is reduced. The fact that $T_2$ is power independent follows from the data. Homogeneous broadening, affecting $T_2$, can be caused by carrier-phonon and carrier-carrier interactions that take place within the exciton lifetime. Carrier-phonon interactions depend on temperature, as is clearly observed in Fig. 2b, the homogeneous coherence time increases with reduced temperature. In contrast carrier-carrier interactions depend on the power. But since the carriers are decaying quickly or trapped within picosecond timescales as compared to the exciton lifetime of 1.6\,ns, then the exciton is free from homogeneous broadening through carrier-carrier interactions for 99.9\,$\%$ of its lifetime.

We point out that the observation of power narrowing for the single quantum dot emission spectrum is not limited to 300\,mK, but is also observed at temperatures up to 12\,K (see Supplementary Information). This outcome suggests that the activation energy of charge traps in the nanowire is above 12\,K, consistent with photoluminescence measurements in our previous work. In all of our samples, we observe weak emission associated with donor-acceptor levels, at approximately 50\,meV below the pure wurtzite InP nanowire peak at 1.49\,eV \cite{Dalacu12}.


\section{Monte Carlo simulations of the quantum dot emission spectrum}
\begin{figure*}[t]
\centering
\includegraphics[width=12cm]{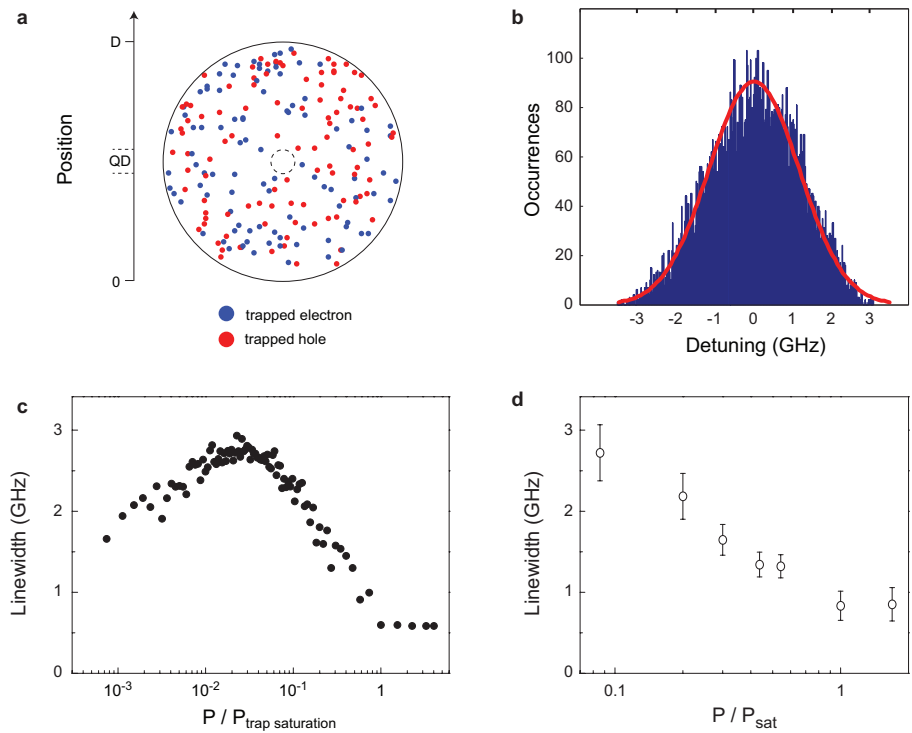}
\caption{\label{fig:4} \textbf{Monte Carlo simulations of Gaussian line broadening mechanism for single quantum dot emission line.} (a) A projection of randomly positioned trap sites for both electrons (blue) and holes (red dots) over a 400\,nm nanowire segment. The nanowire diameter, D, is 200\,nm. The trap density used for the simulations to reproduce the measured exciton linewidth as a function of power is $\sim1\times10^{16}$cm$^{-3}$. The quantum dot, QD, in the center of the nanowire is represented as a dotted circle with diameter of 20\,nm. (b) Resulting histogram of Stark shift induced on the exciton transition for 10,000 possible charge configurations distributed across trap sites surrounding the quantum dot for an average trap occupancy probability is 0.5. Due to the random nature of the possible charge configurations, the broadening mechanism of the emission line is a Gaussian (fit curve in red). (c) Monte Carlo simulations showing the quantum dot emission linewidth as a function of excitation power. The simulations are in excellent agreement with the experimental power dependent emission linewidth presented in (d). P$_{trap\ saturation}$ corresponds to the laser power used to fill trap sites, whereas P$_{sat}$ corresponds to the excitation power used to saturate the charged exciton transition.}
\end{figure*}

We now present Monte Carlo simulations of the quantum dot emission spectrum, which take into account charge fluctuations in the quantum dot environment as a function of power. To reproduce the exciton emission linewidth, we model 1000 randomly positioned electron and hole traps in the nanowire, corresponding to a trap density of $\sim1\times10^{16}$\,cm$^{-3}$. This trap density gives approximately 50 trap sites within a radius of 200\,nm from the dot. Fig.~\ref{fig:4}(a) displays the random positioning of electron (blue) and hole (red) trap sites in the nanowire. In our experiments, the excitation process above the nanowire bandgap excites both the quantum dot and carriers in the nanowire, which subsequently fill trap sites. The interaction between the trapped charges and the quantum dot exciton results in a shifting of the exciton emission energy via the Stark effect. The result of the Monte Carlo simulations iterating over 10000 possible charge configurations distributed across the randomly positioned trap sites for a trap occupation probability of 0.5 is shown in Fig.~\ref{fig:4}(b). Owing to the random nature of the charge fluctuations, the resulting broadening of the emission line is a Gaussian (red fit to the histogram).

Our Monte Carlo simulation gives the emission linewidth as a function of trap occupancy. The correspondence between laser power and trap occupancy follows a non-linear relation and is calculated with a rate equation model (see Supplementary Information). The simulation results of the emission linewidth as a function of excitation power are shown in Fig.~\ref{fig:4}(c). First, we observe that the linewidth broadens with increasing excitation power. This is the typical power broadening regime as is also observed for self-assembled quantum dots.  Next, the broadest emission linewidth is attained, corresponding to the situation where half of the trap sites are occupied, thereby providing the highest number of possible charge configurations in the quantum dot environment. Here, the system reaches the highest entropy. By further increasing the excitation power, the power narrowing regime is reached corresponding to a decrease in the emission linewidth of 2.5\,GHz/decade. In this power narrowing regime, the higher number of traps sites that are occupied results in a reduction of the disorder in the quantum dot environment. The linewidth of the dot may therefore be used to probe the entropy of the system. Finally, once nearly all of the charge traps are occupied at higher power no further improvement of the emission linewidth is expected. The simulation results are in excellent agreement with the observed emission linewidth decrease of 1.9\,GHz per decade with increasing power, including the eventual saturation of this reduction, as presented in Fig.~\ref{fig:4}(d).

It is worth mentioning why we reach the power narrowing regime for the first time. We conceptualize that the main reason is due to a unique combination of both the nanowire geometry and reduced trap density in ultraclean wurtzite InP nanowires. Because of these two features photo-generated charges in the nanowire remain localized around the quantum dot and may be used to saturate nearly all of the surrounding charge traps. In contrast, in a bulk sample, saturating surrounding trap sites is extremely difficult owing to a significant increase in available material. Thus, for a bulk sample, carriers may diffuse to trap sites located further away from the quantum dot. This outcome regrettably leaves self-assembled quantum dots currently in the power broadening regime. However, careful design of the photonic structure in order to reduce the surrounding material, such as etching photonic nanowires similar to the work of J. Claudon et al.~\cite{Claudon10}, may enable the power narrowing regime to be reached for self-assembled quantum dots.

\section{Conclusion}
We have shown for the first time single-photon coherence measurements from a quantum dot in a nanowire waveguide and we found a homogeneous coherence time of 1.2\,ns at 300\,mK, which is the longest coherence time for a quantum dot reported to date. Importantly, this longest single-photon coherence is not attained at very low excitation powers, but at the excitation power where the quantum dot single-photon emission rate is the highest. We overcame power broadening by growing ultraclean wurtzite InP nanowires without stacking faults to reduce the trap density. This reduced trap density combined with the nanowire geometry allowed us to fill, for the first time, nearly all of the trap sites surrounding the quantum dot at higher excitation powers, resulting in a narrower emission spectrum. Finally, the enhanced single-photon coherence, brightness and purity that we demonstrate brings this young field of quantum dots in nanowires amongst the leading candidates for implementations of single-photon sources in future quantum and photonic technologies.

\section{Methods}
\noindent
\textbf{Single-photon interference fringe visibility}
\\
\noindent
The interference fringe visibility is defined as:
\begin{equation}
V=\frac{I_{\rm max}-I_{\rm min}}{I_{\rm max}+I_{\rm min}},
\end{equation}
where $I_{max}$ and $I_{min}$ correspond to the case of optimal constructive and destructive interference, respectively. We measure a maximum raw interference fringe visibility for the quantum dot of 95\,$\%$, whereas for a coherent laser with 4\,MHz linewidth we obtain a raw interference fringe visibility of 96\,$\%$. In all of the single-photon interference measurements, we present the fringe visibility corrected for the Michelson system response (see Supplementary Information). This correction is needed due to the limited phase stability of the Michelson setup caused by vibrations and small temperature fluctuations.

\noindent
\\\textbf{Relationship between coherence and linewidth}
\\
\noindent
The form of the fringe visibility decay, $g^{(1)}(\tau)$, for coherent light depends on the type of spectral broadening that applies. For light with a Lorentzian lineshape, the decay of the fringe visibility is exponential, where:
\begin{equation}
g^{(1)}(\tau)\sim\exp\left(-{\frac{|\tau|}{T_2}}\right).
\end{equation}

\noindent
The relationship between the full-width half-maximum in frequency, $\Delta f$, and the coherence time, $T_2$, for the Lorentzian lineshape is given by:
\begin{equation}
\Delta f_{L} = \frac{1}{\pi T_2},
\end{equation}
\noindent
whereas for a Gaussian lineshape, the fringe visibility decay follows the form:
\begin{equation}
g^{(1)}(\tau)\sim\exp\left[-{\frac{\pi}{2}\left(\frac{\tau}{\tau_c}\right)^2}\right].
\end{equation}
In this case, the full-width half-maximum for a Gaussian is:
\begin{equation}
\Delta f_{G}\sim\frac{\sqrt{2ln2}}{\sqrt{\pi}\tau_c},
\end{equation}
\noindent
where $\tau_c$ is the coherence time for the Gaussian component. The decay of the fringe visibility were fit for all data using a Voigt profile. In the time domain the Voigt profile is a product of a Gaussian and Lorentzian:
\begin{equation}
g^{(1)}(\tau) \sim \exp\left[-\frac{\pi}{2}\left(\frac{\tau}{\tau_c}\right)^2-\frac{|\tau|}{T_2}\right].
\end{equation}
\noindent
In the frequency domain the full-width half-maximum of the resulting Voigt profile is \cite{Wurzburg14}:
\begin{equation}
\Delta f_{V}=0.535\Delta f_{L}+\sqrt{0.217 \Delta f_L^2+\Delta f_G^2}.
\end{equation}

\section{Acknowledgements}
We acknowledge K.D. J\"{o}ns for technical support and scientific discussions, as well as T. Braun, S. H\"ofling, and M. Kamp for the cross-correlations measurements presented in the Supplementary Information that were performed during the scientific visit of M.E.R in W\"urzburg. This work was supported by the European Union Seventh Framework Programme 209 (FP7/2007-2013) under Grant Agreement No. 601126 210 (HANAS) and the Dutch Organization for Fundamental Research on Matter (FOM).

\section{Author contributions}
M.E.R, R.W.H, and V.Z designed and conceived the project. M.E.R, and G.B performed the experiments with R.W.H providing software expertise for the single-photon interference measurements. M.E.R. analyzed the data with input from G.B, B.J.W., M.A.M.V, and V.Z. G.B. performed the Monte Carlo simulations with input from M.E.R, B.J.W., M.A.M.V, and V.Z. D.D and P.J.P carried out the nanowire quantum dot growth. D.D and J.L performed the processing required prior to the site-selected nanowire quantum dot growth. M.E.R. led the project under supervision of V.Z. M.E.R wrote the paper with input from all authors. All authors discussed the results and commented on the manuscript.



\newpage
\beginsupplement

{\Large\centerline{\textbf{Supplementary Information}}}

\title{Overcoming power broadening of the quantum dot emission \\in a pure wurtzite nanowire}

\author{M.E. Reimer$^*$, G. Bulgarini, R.W. Heeres, B.J. Witek, M.A.M. Versteegh, and V. Zwiller}
 \email{m.e.reimer@tudelft.nl}
 \affiliation{Kavli Institute of Nanoscience, Technical University of Delft, Delft, The Netherlands}

\author{D. Dalacu, J. Lapointe and P.J. Poole}%
\affiliation{National Research Council of Canada, Ottawa, Canada}

\maketitle

\section{Michelson system response}

To correct for the Michelson system response in all of our single-photon interference measurements, we measured a coherent Matisse laser with 4\,MHz linewidth. For a time delay of 1.2\,ns between both paths of a perfect Michelson interferometer we would see a flat response with fringe visibility of 1. Instead, we observe a decay of the fringe visibility that follows a Gaussian response, which is attributed to a limited phase stability of our Michelson interferometer caused by vibrations and small temperature fluctuations. To correct the raw fringe visibility decay, we divide the data by the Michelson response (Matisse laser) and obtain the corrected data (blue circles). We note that this correction of the system response to the data has little effect on the resulting emission linewidth. From a fit to the fringe visibility decay at 300\,mK, we obtain an emission linewidth of 940(130)\,MHz for the raw data and 880(130)\,MHz for the corrected data.

\begin{figure}[h]
\includegraphics[width=83mm]{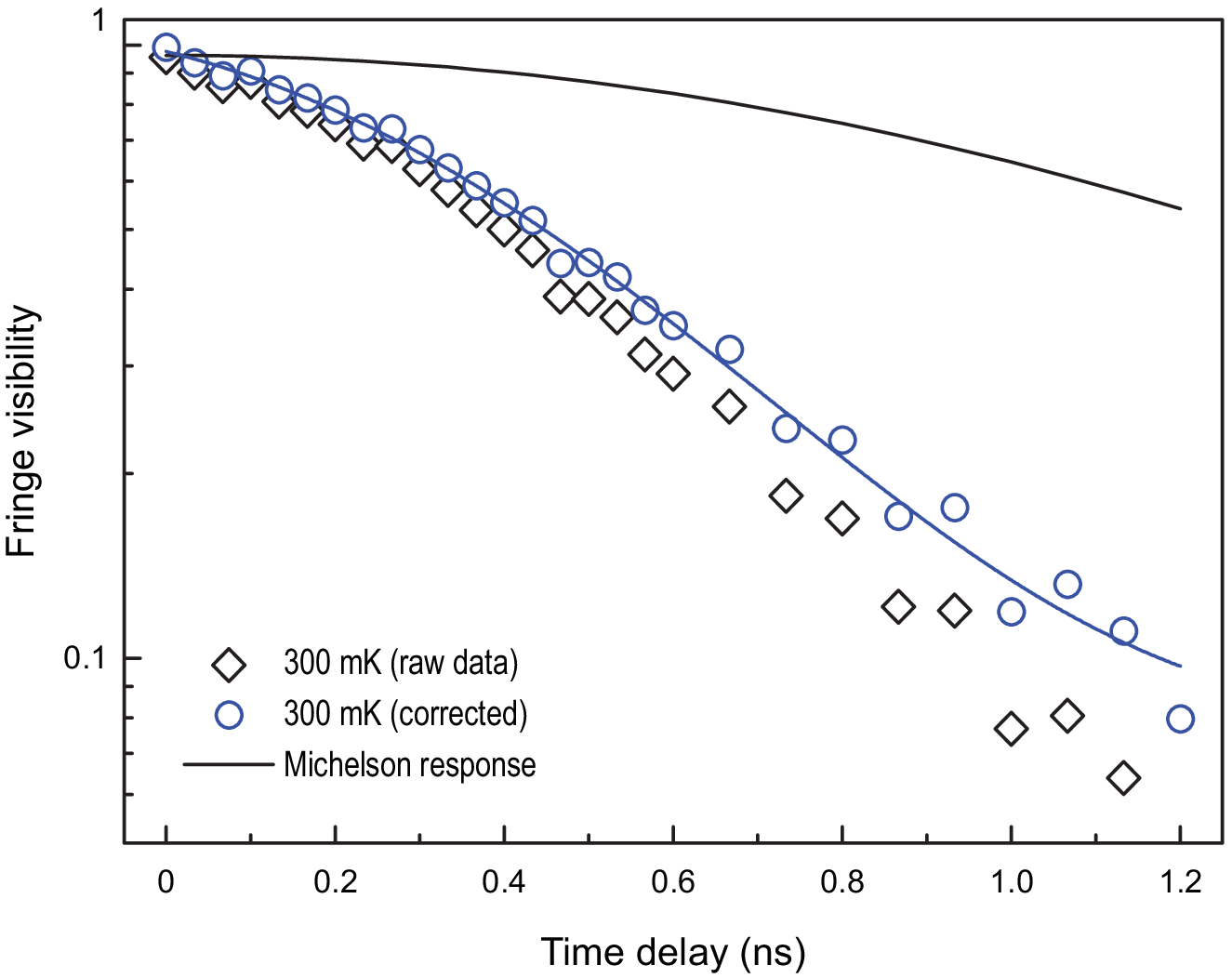}
\caption{Michelson system response. Raw fringe visibility decay at 300\,mK (black diamonds). Michelson response (black line) measured for a coherent Matisse laser with a linewidth of 4\,MHz. On our measured time-scales we expect to observe a flat system response with fringe visibility of 1. Correcting for the system response (raw data/Michelson response), we obtain the corrected data (blue circles).}
\end{figure}

\newpage
\section{Quantum dot exciton lifetime}

\begin{figure}[b]
\includegraphics[width=83mm]{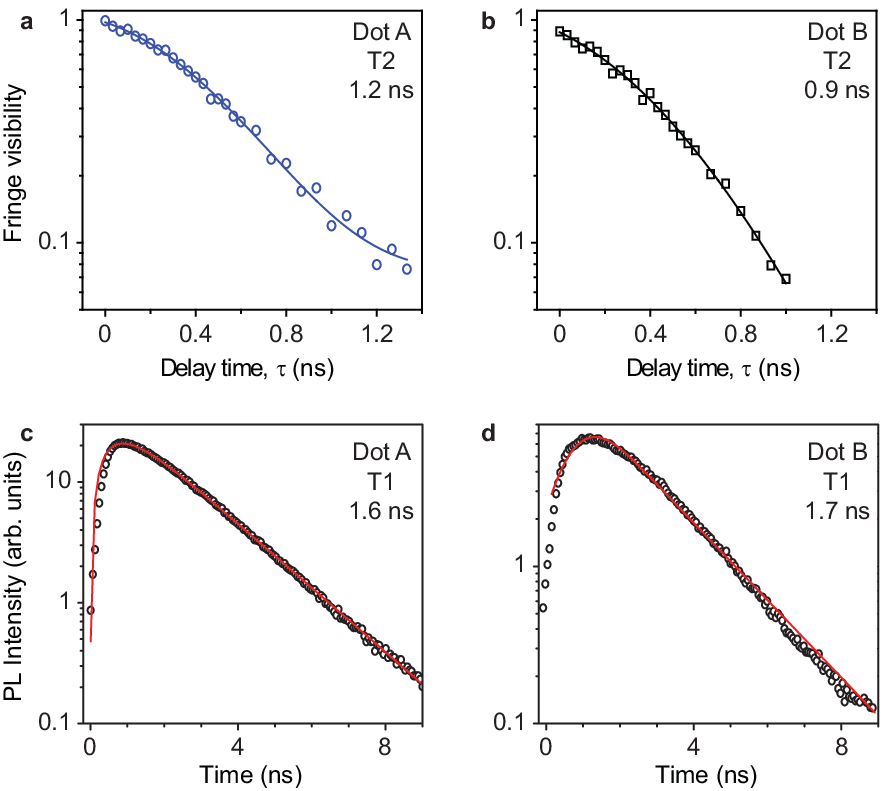}
\caption{\label{fig:supp_Fig7} Single-photon coherence measurements of the charged exciton line of dot A and dot B as compared to the excited state lifetime, $T_1$, at 300\,mK. Dot A is the quantum dot presented in the main manuscript. Single-photon interference measurements of dot A in (a) and dot B in (b). The coherence time, $T_2$, of dot A and B is 1.2(2)\,ns and 0.9(1)\,ns, respectively. The corresponding exciton lifetime measurements of dot A and dot B are shown in (c) and (d), respectively. A fit to the data (orange solid line) yields a $T_1$ for dot A of 1.6(1)\,ns and $T_1$ for dot B of 1.7(1)\,ns. The fit was performed by convoluting the exponential lifetime decay with a timing jitter of 300\,ps for the single-photon avalanche photodiode.}
\end{figure}

We compare the excited state lifetime, $T_1$, to the coherence time, $T_2$, where the theoretical upper limit for a Fourier-transform limited photon is $T_2=2T_1$. The excited state lifetime is measured through time-resolved photoluminescence by exciting the quantum dot with 3\,ps laser pulses using a repetition rate of 76\,MHz and measuring the signal response using a silicon avalanche photodiode. The excited state lifetime of the charged exciton for dot A is presented in Fig.~\ref{fig:supp_Fig7}(c). From a fit to the data, an excited state lifetime, $T_1$, of 1.6(1)\,ns is attained. In comparison to the measured coherence time of 1.2(2)\,ns in Fig.~\ref{fig:supp_Fig7}(a) for dot A, the lifetime Fourier transform limit for this dot is off by a factor of $2T_1/T_2=2.7$. In Fig.~\ref{fig:supp_Fig7}(b), we present the coherence time of another dot (dot B). For dot B, a coherence time of 0.9(1)\,ns is attained. The corresponding excited state lifetime is shown in Fig.~\ref{fig:supp_Fig7}(b), where a $T_1$ of 1.7(1)\,ns is extracted resulting in a ratio of $2T_1/T_2=3.8$. The resulting emission linewidths, including Gaussian line broadening, for dot A and dot B are 880(130)\,MHz and 860(160)\,MHz, respectively.

\section{Identification of charged exciton complexes}

\begin{figure}[b]
\includegraphics[width=83mm]{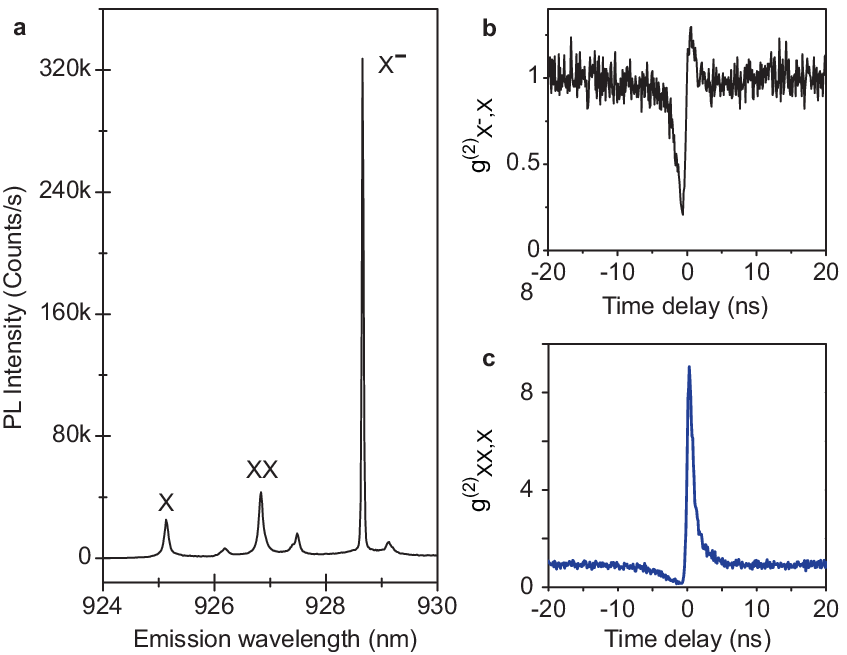}
\caption{\label{fig:supp_Fig9} Identification of charged exciton complexes. (a) Photoluminescence spectrum of quantum dot B that is predominantly charged. (b) Cross-correlation trace of charged (X$^-$) and neutral exciton (X) line. Start: X$^-$; Stop: X. The small bunching at positive time delay indicates that there is a small probability to capture a single hole in the dot prior to re-excitation of the charged exciton state. (c) For comparison the cross-correlation trace of the biexciton-exciton (XX-X) cascade is shown.  Start: XX; Stop: X. The very high bunching indicates a very clean biexciton-exciton cascade.}
\end{figure}

The PL spectrum of quantum dot B used for the field-correlation measurements is presented in Fig.~\ref{fig:supp_Fig9}(b). The brightest line for this particular quantum dot is the singly charged exciton, X$^{-}$. Assignment of the excitonic complexes were confirmed by a combination of both polarization analysis and cross-correlation measurements. In Fig.~\ref{fig:supp_Fig9}(b) we show the cross-correlation trace for the charged (X$^{-}$) and neutral (X) exciton line. In contrast to the typical XX-X cascade that is obtained in Fig.~\ref{fig:supp_Fig9}(c), a diminished bunching peak is observed at small positive time delay. This observed behavior suggests that there is a small probability to recapture a single hole in the dot prior to re-excitation of the charged exciton state. We note that for negative time delay near zero, the signal is anti-bunched since the final state of the exciton is an empty dot and cannot emit another photon until re-excited. For comparison, the cross-correlation measurements of the exciton and biexciton are presented in Fig.~\ref{fig:supp_Fig9}(c). The very high bunching at zero time delay indicates a very clean biexciton-exciton cascade where the exciton photon is immediately emitted after the biexciton photon with a very high probability.
\begin{figure}[h]
\includegraphics[width=83mm]{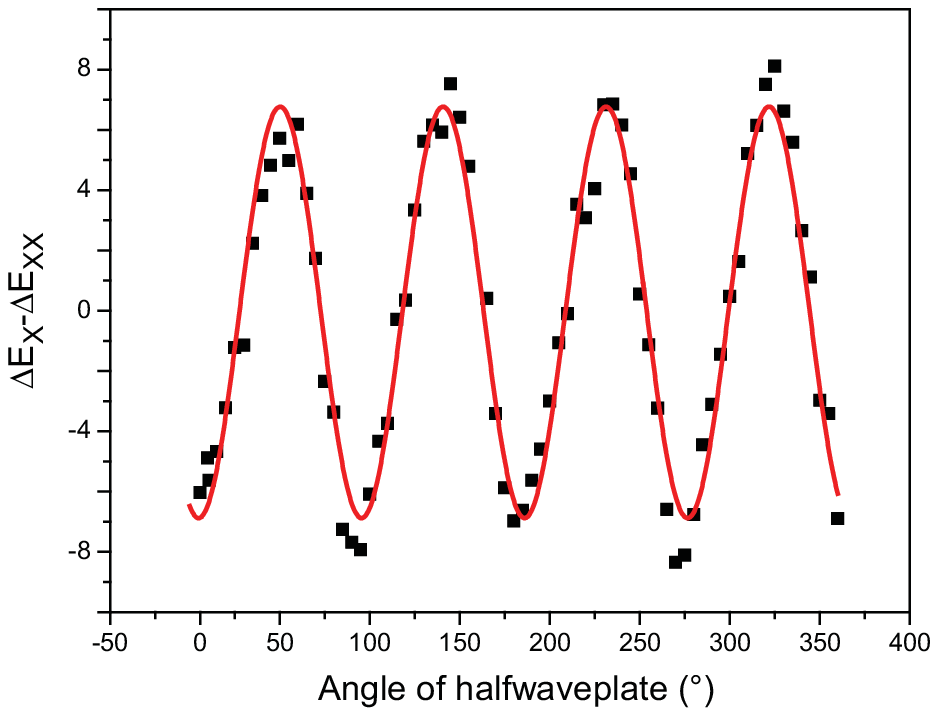}
\caption{\label{fig:supp_Fig10} Polarization dependence of quantum dot B, yielding a fine-structure splitting of 7(1)\,$\mu$eV. We note that the biexciton binding energy has been subtracted away in the presented data.}
\end{figure}

The assignment of exciton complexes are corroborated by polarization analysis. We plot in Fig.~\ref{fig:supp_Fig10} the difference in the peak energies of the exciton, X, and biexciton, XX, for dot B as a function of half-wave plate angle. The amplitude of a sine curve fit to the data in Fig.~\ref{fig:supp_Fig10} yields a very small
splitting of 7(1)\,$\mu$eV. We typically observe a fine-structure splitting for the exciton and biexciton lines between 5 to 10 $\mu$eV across the sample. In contrast, no fine-structure splitting is observed for the charged exciton. This latter lack of fine-structure splitting combined with the observed energetic position of the charged exciton complex at $\sim 5$\,meV below $X$ allows us to assign the charge state as $X^-$. Typically, the negatively charged exciton, $X^-$, is found at lower energy than the neutral exciton, as experimentally verified in our InAsP/InP material system where electrostatic gating enabled us to unambiguously assign the charge state \cite{vanKouwen10}. The exact energy splitting is strongly dependent on the dot morphology \cite{Ediger07} and may be as large as 5\,meV as is typically reported in literature \cite{app_MichaelStark}.

\newpage
\section{Temperature dependence of the quantum dot emission spectrum}
\begin{figure}[h]
\includegraphics[width=83mm]{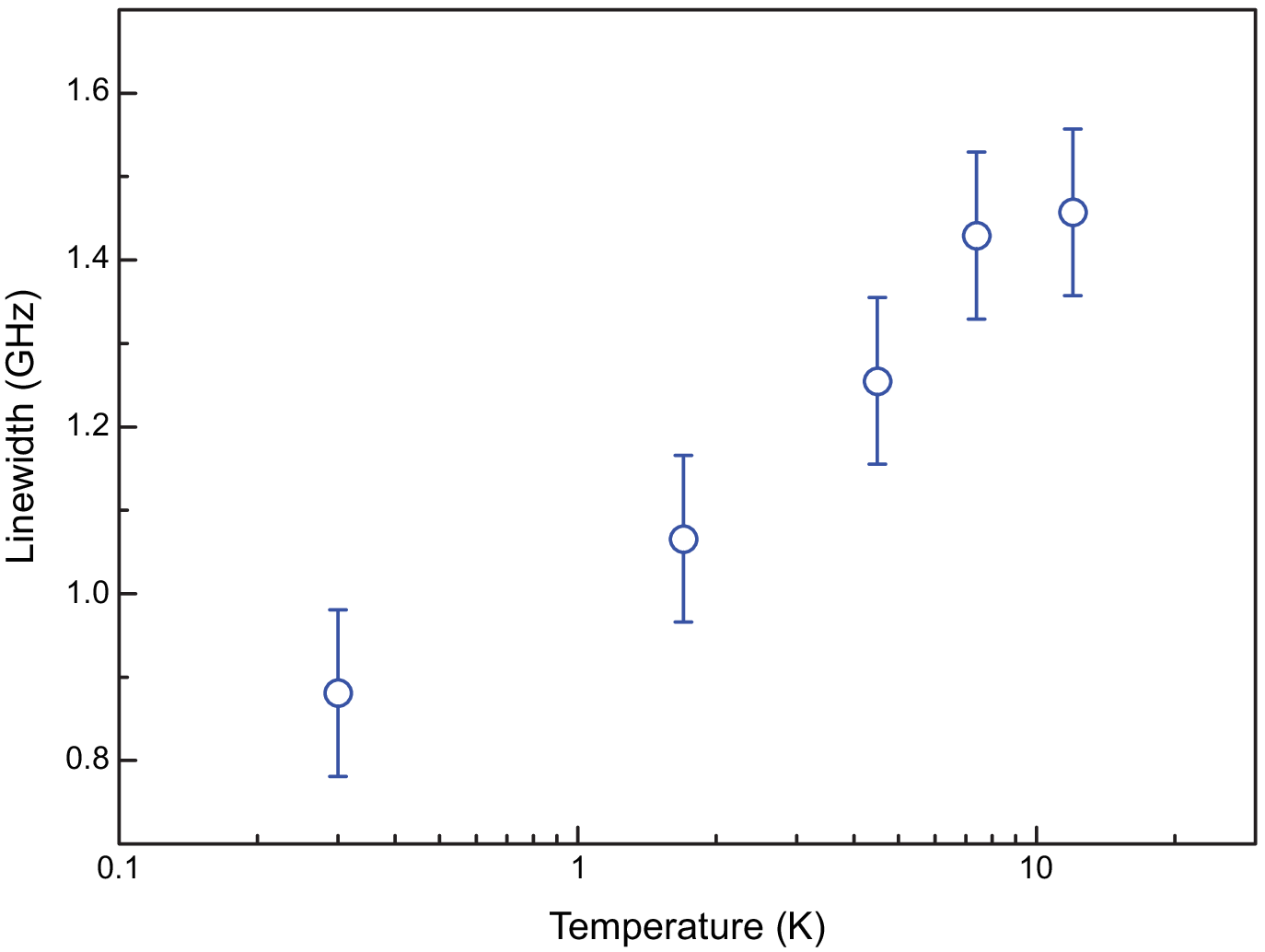}
\centering
\caption{Measured charged exciton linewidth of quantum dot A as a function of temperature at exciton saturation power. The observed linewidth dependence is super-linear and narrows as the temperature is cooled from 12\,K to 300\,mK by suppressing interactions with acoustic phonons.}
\end{figure}

\section{Power narrowing of the quantum dot emission spectrum at higher temperatures}
\begin{figure}[h]
\includegraphics[width=83mm]{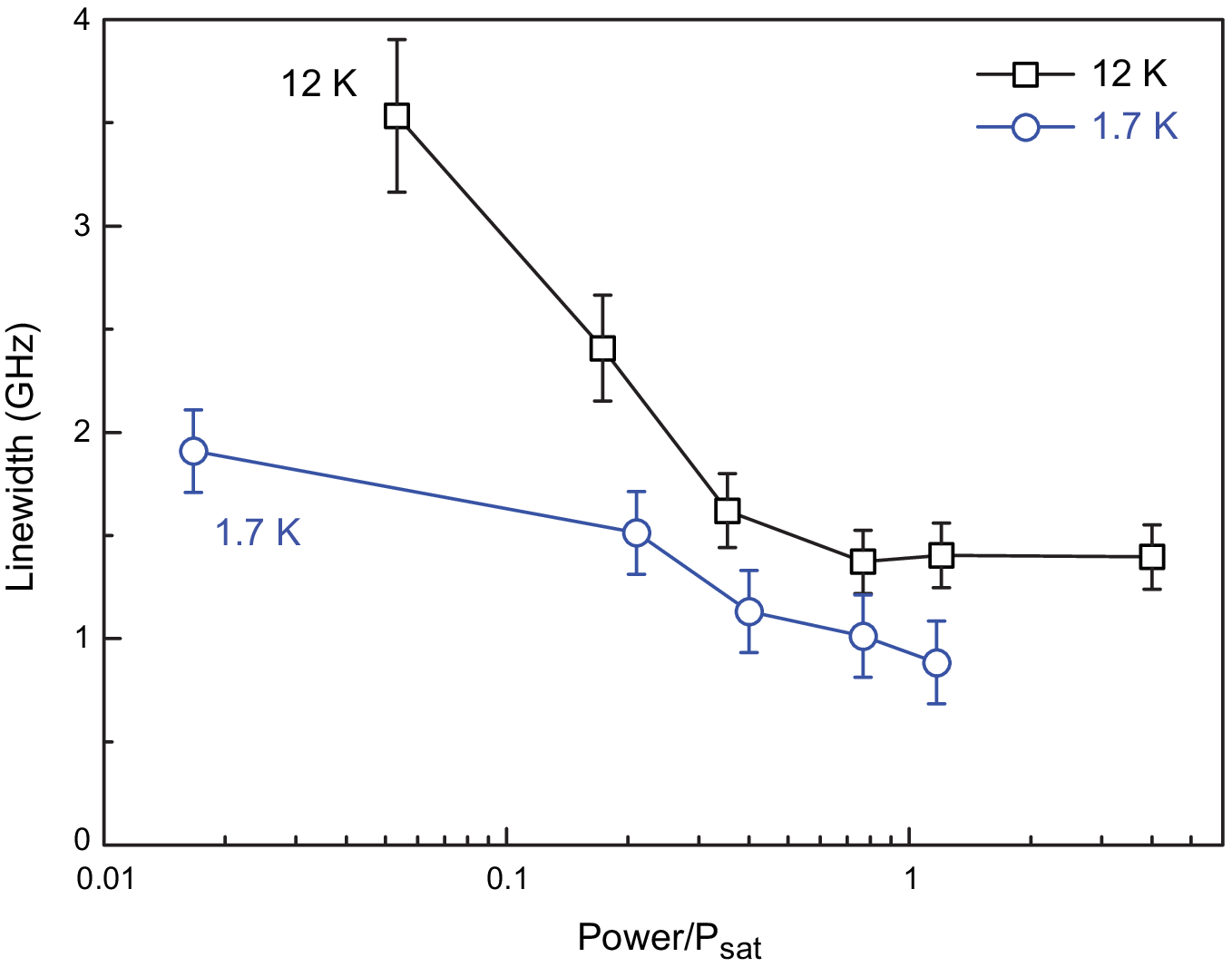}
\centering
\caption{Measured charged exciton linewidth as a function of excitation power for temperatures of 1.7\,K and 12\,K. The power narrowing effect is observed at both temperatures indicating that the activation energy of charge traps in the nanowire is above 12\,K.}
\end{figure}

\section{Single-photon coherence as a function of excitation energy}
\begin{figure}[h]
\includegraphics[width=83mm]{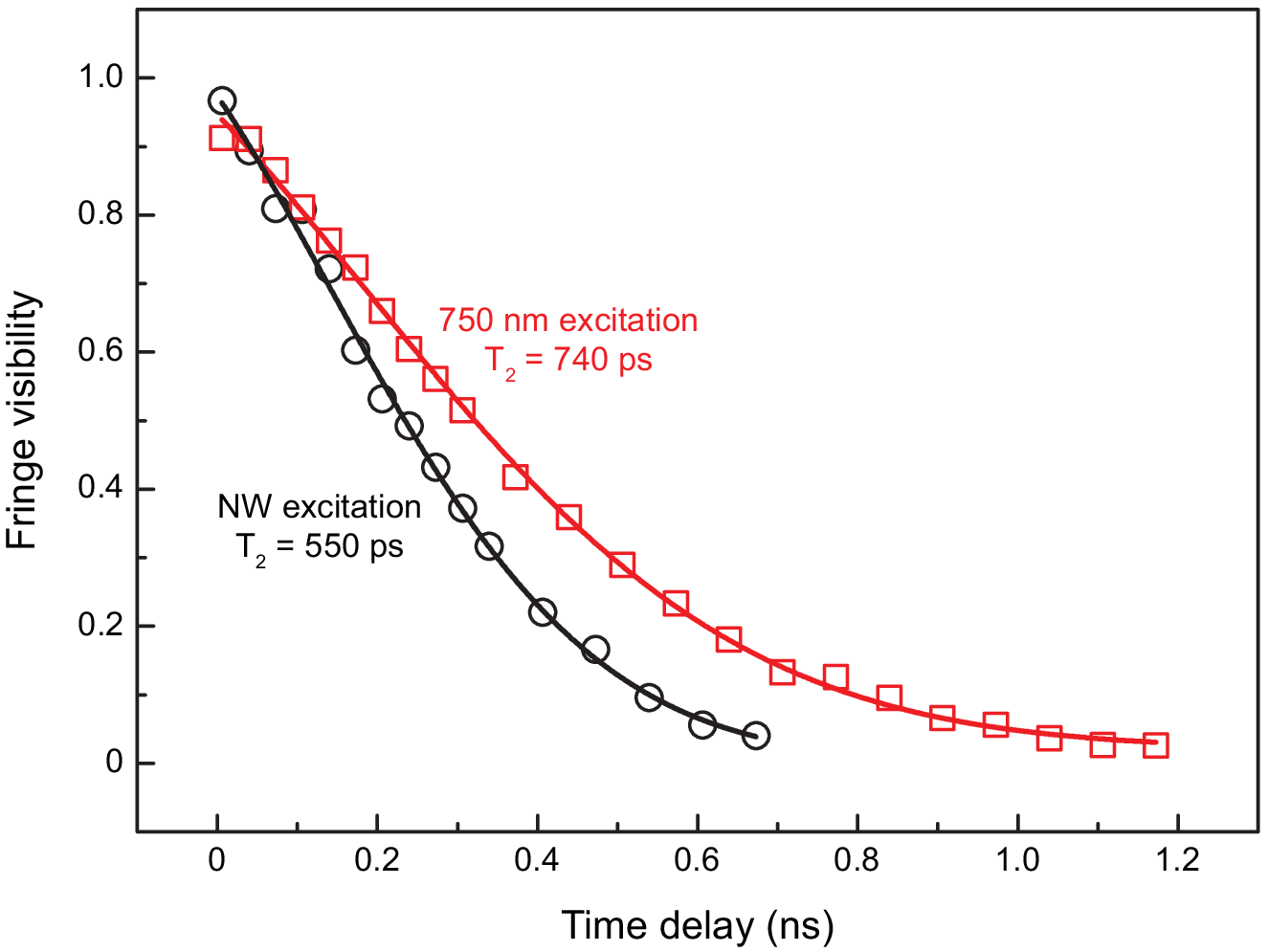}
\centering
\caption{Measured charged exciton linewidth at 1.7\,K for two excitation energies: at the the wurtzite nanowire bandgap (836\,nm) and above the nanowire bandgap (750\,nm). Both measurements were performed at exciton saturation where the single-photon emission rate is highest. The single-photon coherence is reduced at a laser excitation energy of the wurtzite nanowire bandgap, indicating that the surrounding charge traps are not saturated as is the case for higher energy of the excitation laser.}
\end{figure}

\newpage
\section{Monte Carlo simulation of the quantum dot emission spectrum}

\subsection*{Simulation of the quantum dot emission spectrum}
Here we present details of our Monte Carlo simulation model that explains the broadening mechanism of a quantum dot exciton emission line due to fluctuating charges in its surrounding environment. Our model accounts for both the power broadening regime as observed in previous work and the power narrowing regime as demonstrated in this work. With our model, we account for the dynamics of charges randomly distributed in trap sites surrounding the quantum dot in order to evaluate the statistical distribution of emitted single photons. This model describes a quantum dot in a nanowire, but is easily extendable to other geometries.

\subsubsection{Rate equations}

\noindent The model accounts for the line broadening of the quantum dot exciton emission due to the interaction with charges in the environment. Electrons and holes are photo-generated by the above-bandgap excitation laser in the nanowire. Subsequently, those charges immediately separate and are captured into trap sites in the nanowire at a rate $\Gamma_g$, on a time-scale that is faster than the exciton lifetime in the quantum dot. The interaction between the trapped charges and the quantum dot exciton results in a detuning of the exciton emission energy. Trapped charges have a finite probability of recombination at a recombination rate $\Gamma_r$ or a trapped charge may overcome the trapping potential and escape the trap site with an escape rate $\Gamma_e$. We employed the following rate equation in order to obtain a relation between the density of occupied charge traps in the quantum dot environment and the laser excitation power,
\begin{equation}
{\frac{dm}{dt}} \,=\,(N-m)\Gamma_g \,- m \Gamma_e - m \Gamma_r ,
\label{eq1}
\end{equation}
\noindent where \emph{m} is the density of occupied traps and \emph{N} is the total density of traps. The density of occupied trap sites increase at a rate $\Gamma_g$ and it is proportional to the density of free trap sites, \emph{N-m},
\begin{equation}
\Gamma_g = \frac{\eta}{N} \cdot \frac{P}{h\nu \cdot A\cdot L}.
\label{eq2}
\end{equation}
The trap generation rate, $\Gamma_g$ is inversely proportional to the total number of traps and proportional to $\eta$, which represents the trapping probability for a photo-generated charge. Moreover, the trap generation rate is proportional to the ratio between excitation power, $P$, and photon energy $h\nu$, spot area $A$ and sample thickness $L$, under the assumption of unity absorption efficiency used for simplicity. \par
We introduce here the excitation photon flux that is the number of photons per unit time and unit area impinging on the sample. The photon flux is directly proportional to the excitation power:
\begin{equation}
\Phi = \frac{P}{(h\nu \cdot A)}.
\label{eq3}
\end{equation}
Two mechanisms are responsible to reduce the density of occupied traps: escape and recombination. At low temperatures such as a few Kelvin, we assume that thermally activated escape is suppressed. The escape rate is thus proportional to excitation photon flux with a probability, \emph{x}, that a charge escapes from the trap site because re-excited by the laser:
\begin{equation}
\Gamma_e=\frac{x\Phi}{N \cdot L}.
\end{equation}
\noindent By combining \ref{eq2} and \ref{eq3}, equation \ref{eq1} can be written in the following form:
\begin{equation}
{\frac{dm}{dt}} \,=\,(N-m) \frac{\eta}{L} \frac{\Phi}{N} \,- m \frac{x}{L}\frac{\Phi}{N} - m \Gamma_r ,
\end{equation}
\noindent
At equilibrium we have a steady number of trapped charges,
\begin{equation}
\frac{dm}{dt}=0.
\end{equation}
\noindent
Hence, the density of occupied traps in the semiconductor structure is
\begin{equation}
m\,=\,\frac{\eta}{L}\cdot\frac{\Phi}{\Phi(\frac{\eta + x}{L})/N+\Gamma_r}.
\label{eqmN}
\end{equation}

\begin{figure*}[t]
\includegraphics[width=0.9\textwidth]{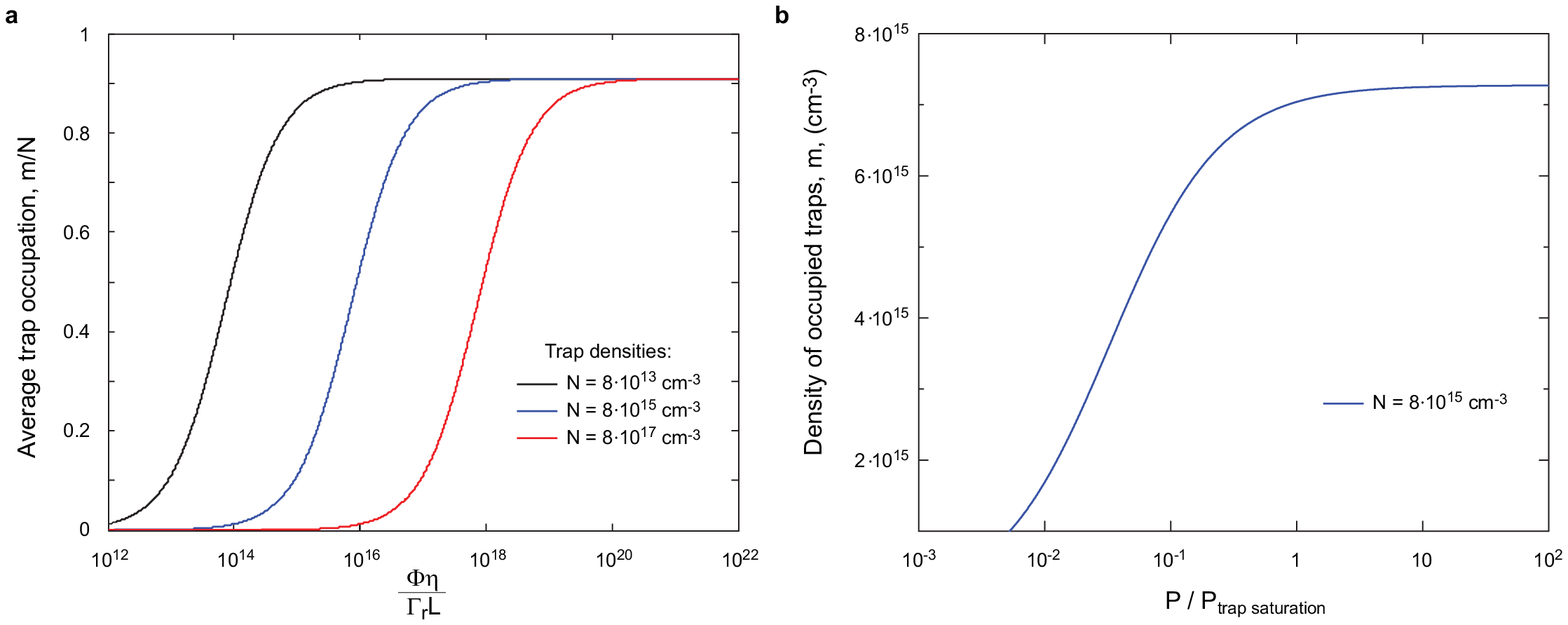}
\centering
\caption{(a) Average trap occupation for trap density, \emph{N}, equal to 8\,$\times$\,$10^{13}$\,cm$^{-3}$ (black), 8\,$\times$\,$10^{15}$\,cm$^{-3}$ (blue) and 8\,$\times$\,$10^{17}$\,cm$^{-3}$ (red). (b) Density of occupied traps, m, as a function of the excitation power, \emph{P}, calculated using a trap density 8\,$\times$\,$10^{15}$\,cm$^{-3}$ from (a).}
\label{fig:app1}
\end{figure*}
\noindent Fig. \ref{fig:app1} shows the average trap occupation for different trap densities as a function of excitation photon flux. The probability of laser-induced escape process is assumed to be 10\,$\%$ as compared to the trapping probability. This sets the occupation saturation at 90\,$\%$ of the total number of traps. We observe that a higher density of traps implies a higher power needed to saturate their occupation. This excitation power may extend beyond the power used for saturating the exciton population in the quantum dot.

\subsubsection{Monte Carlo simulations}
\noindent
The interaction between fluctuating trapped charges on the exciton emission line is calculated using Monte Carlo simulations. We first generate randomly distributed trap sites for electrons and holes in the nanowire. The nanowire is a cylinder of 200\,nm in diameter and 4\,$\mu$m in length. No trap sites are placed in the quantum dot. These traps are randomly occupied with a probability set by the average trap occupation, which is calculated via the rate equation model presented in the previous section. A shift of the exciton emission energy arises from interaction with the charge environment and is determined via the Stark effect. The electric field generated from a trapped charge and quantum dot exciton is calculated in the single particle picture,
\begin{equation}
F\,=\frac{\,e}{4\pi \epsilon r^2},
\end{equation}
where $\epsilon$ is the dielectric constant of the semiconductor matrix and \emph{r} is the distance between the exciton and the trapped charge. The Stark shift of the exciton subject to an electric field, $F$, is given by,
\begin{equation}
V\,=\,\beta\,F^2,
\end{equation}
where $\beta$ is a material dependent constant. The Stark shift is thus proportional to 1/${r^4}$. For our material system, InAsP quantum dot embedded in InP, we used the value of $\beta$ calculated for an InAs quantum dot in InP\cite{app_MichaelStark} ($\beta = 1.7 \mu$eV$\cdot$cm$^2$/kV$^2$) and the dielectric constant of InP, $\epsilon = 12.4$.\\
\begin{figure*}[t]
\centering
\includegraphics[width=0.7\textwidth]{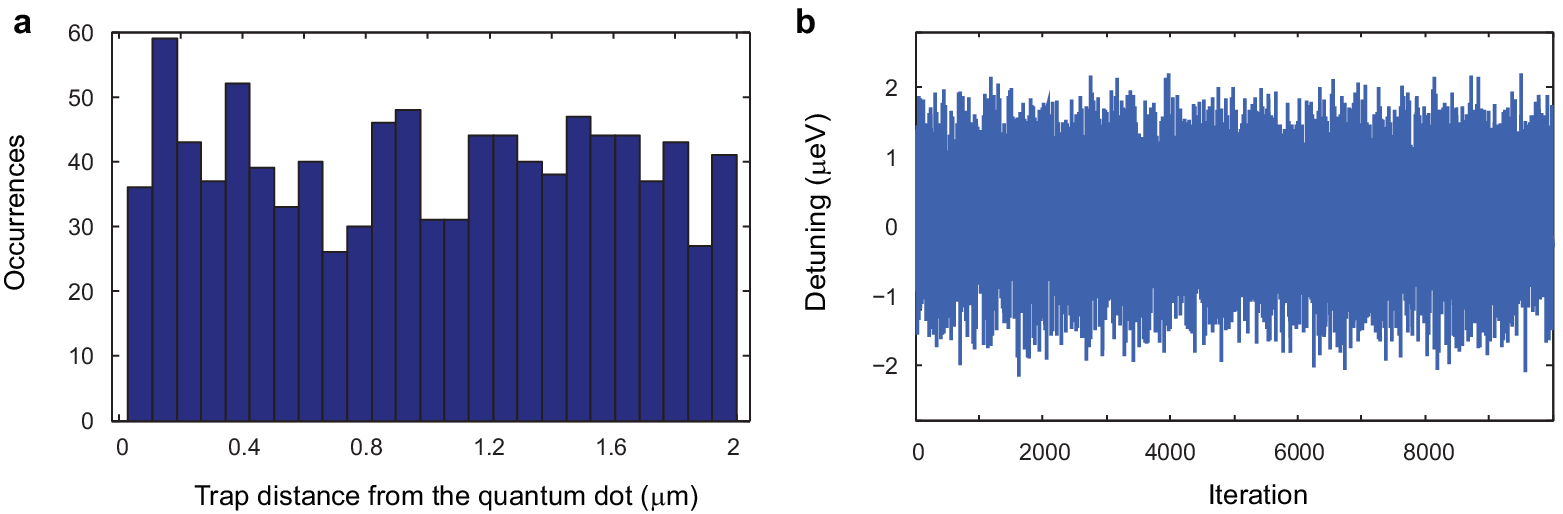}
\caption{(a) Distribution of the trap distance from the quantum dot for 1000 electron traps. (b) Stark shift induced on the exciton transition for each possible charge configuration. We consider 1000 electron traps and 1000 hole traps at an average trap occupancy of 0.5.}
\label{fig:app2}
\end{figure*}
\noindent
\par
\noindent In the example of Fig. \ref{fig:app2}(a), we show the statistical distribution of 1000 trapped charges as a function of the distance to the quantum dot center. We calculate the Stark shift for 10000 randomly selected configurations of electrons and holes in the traps. As an example, Fig. \ref{fig:app2}(b) shows a simulation for an average trap occupancy of $m/N\,=\,0.5$. The statistical distribution of the Stark shift is plotted in the histogram of Fig. 4(b) in the main text.

\subsection*{Power dependence of the emission linewidth}
\begin{figure}
\centering
\includegraphics[width=83mm]{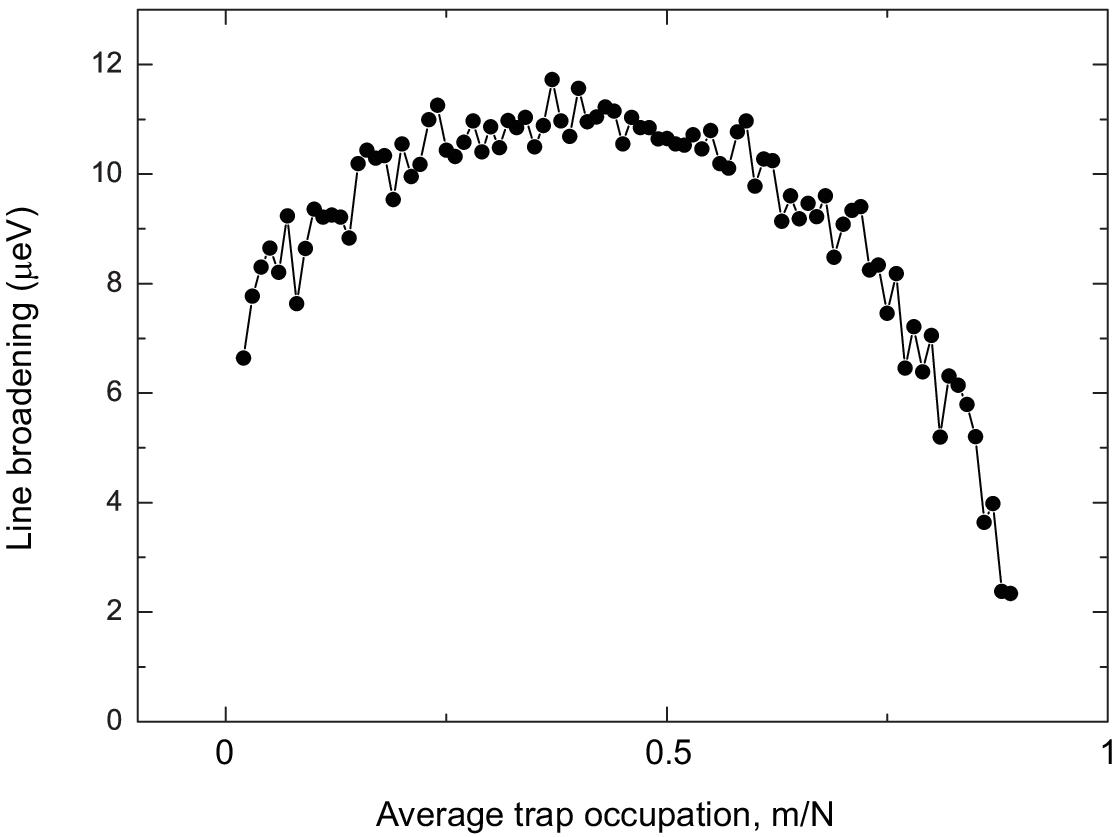}
\caption{Result of Monte Carlo simulations of the exciton line broadening for 1000 electron traps and 1000 hole traps, corresponding to a density of 8\,$\times$\,$10^{15}$\,cm$^{-3}$, as a function of average trap occupation.}
\label{fig:app4}
\end{figure}
\noindent We now calculate the exciton emission linewidth dependence as a function of the average trap occupation, $m/N$. In this further step, we include the possibility for the quantum dot environment of being n-doped as InP nanowires generally have a background doping that is n-type. We choose for the simulation, shown in Fig. \ref{fig:app4}, a concentration of 8\,$\times$\,$10^{14}$\,cm$^{-3}$ for n-dopants. The result of the excess electron is to increase the occupation for electron traps as compared to the non-doped case. The results of the Monte Carlo simulations are shown in Fig. \ref{fig:app4}. We observe that the exciton transition linewidth broadens while increasing the average trap occupation from 0 to 0.5, where the line broadening is maximum. Here, the increased interaction with the fluctuating charge environment broadens the energy spectrum of the emitted photons. The maximum multiplicity of charge configurations in the environment is thus found for a trap occupation probability of 0.5. If more than half of the traps are filled the system goes towards a more stable charge environment: the number of possible charge configurations is limited and the emission linewidth decreases. Eventually, when all the traps are filled, the charge environment of the quantum dot would be stationary. However, this condition is never fulfilled because of the finite de-trapping of charges via laser absorption. The background doping of the quantum dot environment gives a non-zero broadening of the exciton line even at very low excitation power. Furthermore, depending on the specific position of the randomly located charges, the line broadening varies from simulation to simulation. The maximum line broadening that we obtain at half trap occupation ranges from 6 $\mu$eV to 11 $\mu$eV. This random positioning of nanowire trap sites explains the variance in quantum dot emission linewidth from dot to dot on the same sample. Finally, by combining the rate equations with the Monte Carlo simulations, we obtain the results presented in Fig. 4\textbf{c} of the main text where the emission linewidth dependence is plotted as a function of excitation power.


\end{document}